\documentclass{optica-article}

\journal{opticajournal} 

\articletype{Research Article}

\usepackage{lineno}

\begin{document}

\title{Sensitive Schrödinger Cat States}

\author{Shahab Ramezanpour\authormark{1,*}}

\address{\authormark{1}The Edward S. Rogers Sr. Department of Electrical and Computer Engineering, University of Toronto, 10 King's College Road, Toronto, Ontario, M5S 3G4, Canada}
\email{\authormark{*}shahab.ramezanpour@utoronto.ca} 


\begin{abstract*} 
Strong laser-atom interactions can produce highly non-classical states of light by using the process of high-harmonic generation in atoms. When the high-harmonic generation is present, the quantum state of the fundamental mode following the interaction is known as the Schrödinger cat state, which is a superposition of the laser's initial coherent state and the coherent state with a smaller amplitude that results from its interaction with atoms. Here, we demonstrate that new light states with significantly different Wigner function distributions can be produced by combining two separate Schrödinger cat states. Through the engineering of Schrödinger cat states' parameters, we are able to produce Wigner functions that exhibit high sensitivity in relation to the system parameter. Our research paves the way for the creation of non-classical light by superposing Schrödinger cat states with application in such as quantum sensing.
\end{abstract*}
\section{Introduction}
Numerous well-known methods for producing quantum light at optical frequencies rely on substances with a nonlinear optical response. Such nonlinear materials can often be explained by a "perturbative" nonlinear response, where the induced polarisation in the applied electric field is, for instance, quadratic or cubic.
Non-perturbative or "strong-field" effects, such as high-harmonic generation (HHG), are at the other end of the nonlinear optics spectrum \cite{2}. In HHG, a highly intense optical pulse generates radiation at extremely high frequencies, even more than a hundred times the drive frequency \cite{3}. As a result, HHG is a desirable source of extremely ultra-short high-frequency light pulses. However, the potential of HHG for producing non-classical high-frequency light has largely gone unexplored.

There are various natural science applications for the generation and manipulation of many-photon quantum states of light. Squeezed light \cite{4,5}, Schrödinger kitten \cite{6,1}, and cat states \cite{7,8}, among other many-photon quantum states of light, have all been established by a number of groundbreaking research. In the detection of gravitational waves, for instance, squeezed quantum light states enable extraordinarily sensitive measurements that go beyond the boundaries of classical noise \cite{9,10}.

In addition to fundamental physical interest, revealing the quantum character of light in highly laser-driven interactions is crucial for applications in basic science and technology \cite{1}. This is due to the fact that it fills the gap between strong-laser-field physics \cite{11,12}, and quantum optics \cite{13}, enabling the development of strong-field quantum electrodynamics and a new class of non-classical light sources which are at the foundation of quantum technology \cite{14,15}.

If the starting state of a system of N atoms in their ground state is impinged by a laser in a coherent state with amplitude ${{\alpha }_{L}}$, the quantum states of the fundamental and harmonic modes are coherent. However, the fundamental mode amplitude is shifted as a result of coherence and correlations between the fundamental and the harmonics, (i.e., ${{\alpha }_{L}}\to {{\alpha }_{L}}+\delta {{\alpha }_{L}}$), where $\delta {{\alpha }_{L}}$ is negative and denotes energy conservation; in the context of the semiclassical model, this represents the energy losses of the fundamental mode resulting from the re-collision process. In the experimental setup in ref. \cite{1}, the ultimate state of the fundamental mode, subject to harmonic generation, is a superposition of the initial coherent state ($\left| \alpha_{L}\right\rangle$) and shifted coherent state ($\left| {{\alpha }_{L}}+\delta {{\alpha }_{L}} \right\rangle$) as

\begin{equation}
\left| \alpha_{cat}\right\rangle= \left| {{\alpha }_{L}}+\delta {{\alpha }_{L}} \right\rangle -\zeta\left| {{\alpha }_{L}} \right\rangle  
\end{equation}
, where $\zeta=\left\langle  {{\alpha }_{L}} | {{\alpha }_{L}}+\delta {{\alpha }_{L}} \right\rangle$. 

In this paper we study superposition of two cat states, and reveal that their Wigner function can be highly sensitive versus the parameter space. Its most well-known optical counterpart leveraging nonlinearity or perturbation is perhaps \emph{Exceptional Points} \cite{18,19,20}, a degeneracy in non-Hermitian systems, from which sensitivity can be enhanced.
\section{Superposition of Schrödinger Cat States}
Figure 1 shows that for a specific parameters, the Wigner function of the superposition of cat States is similar to a single cat State, while with slightly changing the parameters, the Wigner function changes significantly.
\begin{figure} 
\centering
{\includegraphics[width=.5\textwidth]{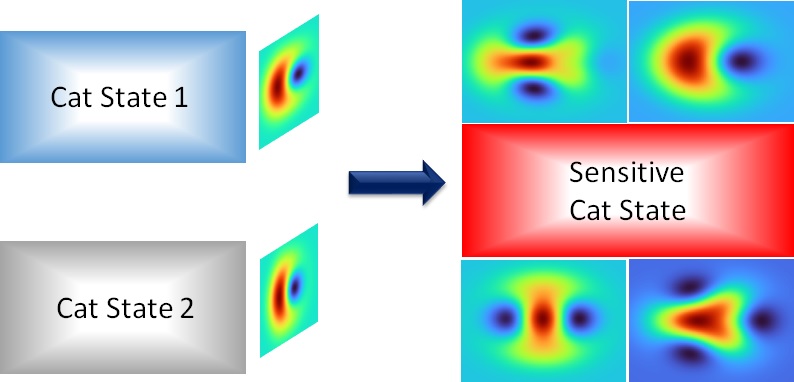}}
\caption{Superposition of cat states can lead to a novel state whose Wigner function changes abruptly in parameter space.} 
\end{figure}
To generate superposition of cat states, we follow the experimental setup of reference \cite{1} but with two blocks with separate laser pulses (Fig. 2). In each block, the Xe gas in upper branch generates higher harmonics while the fundamental laser pulse is carried by lower branch. The output of block one, for instance, is superposition of the initial coherent state $\left| {{\alpha }_{0}} \right\rangle $, and shifted coherent state $\left| {{\alpha }_{1}} \right\rangle =\left| {{\alpha }_{0}}+\delta {{\alpha }_{0}} \right\rangle $, as
$\left| \alpha_{cat,1}  \right\rangle =\left| {{\alpha }_{1}} \right\rangle -\xi_1 \left| {{\alpha }_{0}} \right\rangle $, where $\xi_1 =\left\langle  {{\alpha }_{0}} | {{\alpha }_{1}} \right\rangle $.
\begin{figure} 
\centering
{\includegraphics[width=.5\textwidth]{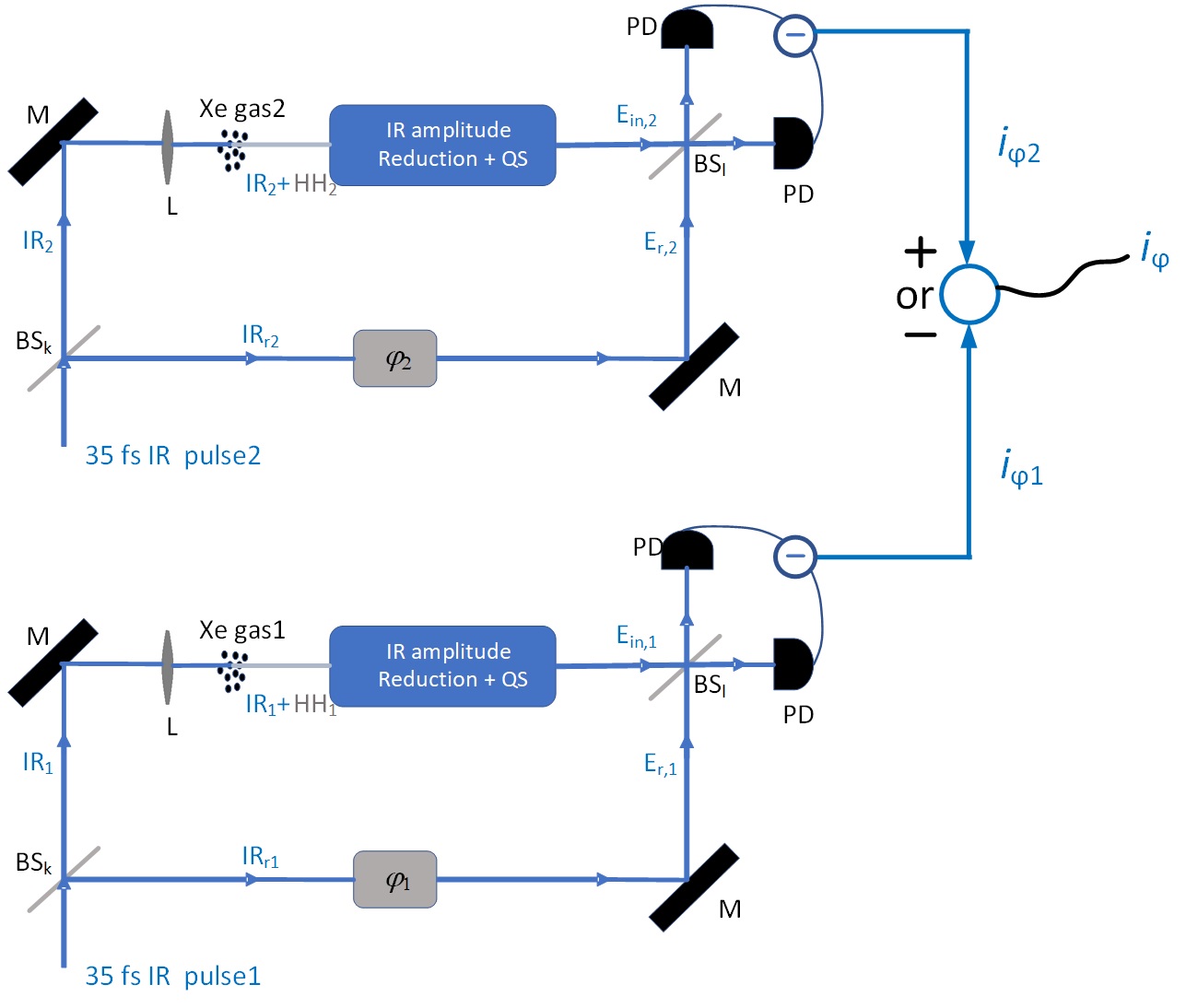}}
\caption{Experimental setup for superposition of Schrödinger cat states. The IR pulse passes through beam splitter ($BS_k$) and mirror (M), and focused by a lens (L) onto a xenon gas jet where high harmonics (HH) is generated, and characterized by quantum spectroscopy (QS) with the output $E_{in}$.  The beam reflected by the $BS_k$, $IR_{r}$,  experiences phase shift $\varphi$. The fields $E_r$ and $E_{in}$ pass through $BS_l$, where two identical photodetectors (PD) are used to generate photocurrent difference ($i_{\varphi1,2}$) to construct Wigner function.}
\end{figure}
The parameters in each of the generated cat state may be tuned by amplitude of the laser pulse and thickness of the Xe gas section. The output light is superposition of two different cat states.  

We investigate superposition of two cat states $\left| {{\alpha }_{cat,1}} \right\rangle=\left| {{\alpha }_{1}} \right\rangle -{{\zeta }_{1}}\left| {{\alpha }_{0}} \right\rangle$ and $\left| {{\alpha }_{cat,2}} \right\rangle=\left| {{\alpha }_{2}} \right\rangle -{{\zeta }_{2}}\left| {{\alpha }_{00}} \right\rangle$ as
\begin{align}
&\left| \alpha_{cat,s}  \right\rangle =\left( \left| {{\alpha }_{1}} \right\rangle -{{\zeta }_{1}}\left| {{\alpha }_{0}} \right\rangle  \right)- \left( \left| {{\alpha }_{2}} \right\rangle -{{\zeta }_{2}}\left| {{\alpha }_{00}} \right\rangle  \right) \\
&\left| \alpha_{cat,s}  \right\rangle =\left( \left| {{\alpha }_{1}} \right\rangle -{{\zeta }_{1}}\left| {{\alpha }_{0}} \right\rangle  \right)+ \left( \left| {{\alpha }_{2}} \right\rangle -{{\zeta }_{2}}\left| {{\alpha }_{00}} \right\rangle  \right)
\end{align}
, where 
\begin{align}
  & \left| {{\alpha }_{1}} \right\rangle =\left| {{\alpha }_{0}}+\delta {{\alpha }_{0}} \right\rangle \quad ;\quad {{\zeta }_{1}}=\left\langle  {{\alpha }_{0}} | {{\alpha }_{1}} \right\rangle  \nonumber\\ 
 & \left| {{\alpha }_{2}} \right\rangle =\left| {{\alpha }_{00}}+\delta {{\alpha }_{00}} \right\rangle \quad ;\quad {{\zeta }_{2}}=\left\langle  {{\alpha }_{00}} | {{\alpha }_{2}} \right\rangle  \nonumber
\end{align}

\subsection{Difference}
As an numerical example, we depict Wigner function of $\left |\alpha_{cat,s}\right\rangle$ in eq. (2), for ${{\alpha }_{0}}=2,{{\alpha }_{00}}=2.3,\ \delta {{\alpha }_{0}}=-1$, and different values of $\delta {{\alpha }_{00}}=-1.2$ (Fig. 3(a)), $\delta {{\alpha }_{0}}=-1.3$ (Fig. 3(b)), $\delta {{\alpha }_{0}}=-1.4$ (Fig. 3(c)), $\delta {{\alpha }_{0}}=-1.5$ (Fig. 3(d)). The Wigner function calculation can be found in the appendix. The figure 3  rapid change in the Wigner function, together with a slight variation around $\delta\alpha_{00}=-1.3$ is one of its most intriguing features. For $\delta\alpha_{00}=-1.3$, $\left |\alpha_1\right\rangle=\left |\alpha_2\right\rangle$, it has the same Wigner function as a single cat state, however, around this point, the Wigner function changes abruptly.
\begin{figure} 
\centering
{\includegraphics[width=.95\textwidth]{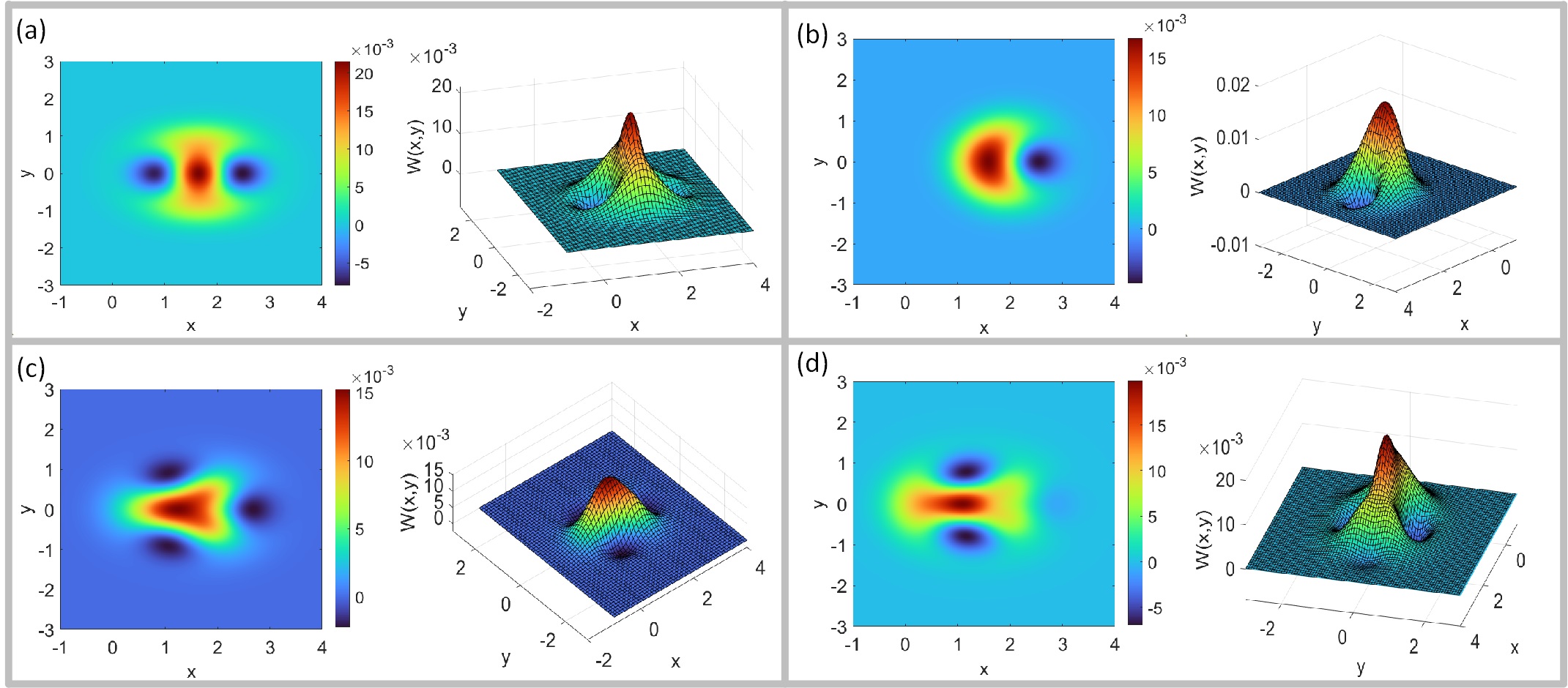}}
\caption{Wigner function for $\left| \alpha_{cat,s}  \right\rangle =\left( \left| {{\alpha }_{1}} \right\rangle -{{\zeta }_{1}}\left| {{\alpha }_{0}} \right\rangle  \right)- \left( \left| {{\alpha }_{2}} \right\rangle -{{\zeta }_{2}}\left| {{\alpha }_{00}} \right\rangle  \right)$ with ${{\alpha }_{0}}=2,{{\alpha }_{00}}=2.3,\ \delta {{\alpha }_{0}}=-1$, and different values of $\delta {{\alpha }_{00}}=-1.2$ (a), $\delta {{\alpha}_{00}}=-1.3$ (b), $\delta {{\alpha }_{00}}=-1.4$ (c), $\delta {{\alpha}_{00}}=-1.5$ (d).}
\end{figure}
\subsection{Sum}
We depict Wigner function of $\left |\alpha_{cat,s}\right\rangle$ in eq. (3) for ${{\alpha }_{0}}=4,{{\alpha }_{00}}=2,\ \delta {{\alpha }_{00}}=-1.5$, and different values of $\delta {{\alpha }_{0}}=0$ (Fig. 4(a)), $\delta {{\alpha }_{0}}=-0.5$ (Fig. 4(b)), $\delta {{\alpha }_{0}}=-1$ (Fig. 4(c)), $\delta {{\alpha }_{0}}=-2$ (Fig. 4(d)), $\delta {{\alpha }_{0}}=-2.5$ (Fig. 4(e)), $\delta {{\alpha }_{0}}=-3$ (Fig. 4(f)). 
\begin{figure} 
\centering
{\includegraphics[width=.95\textwidth]{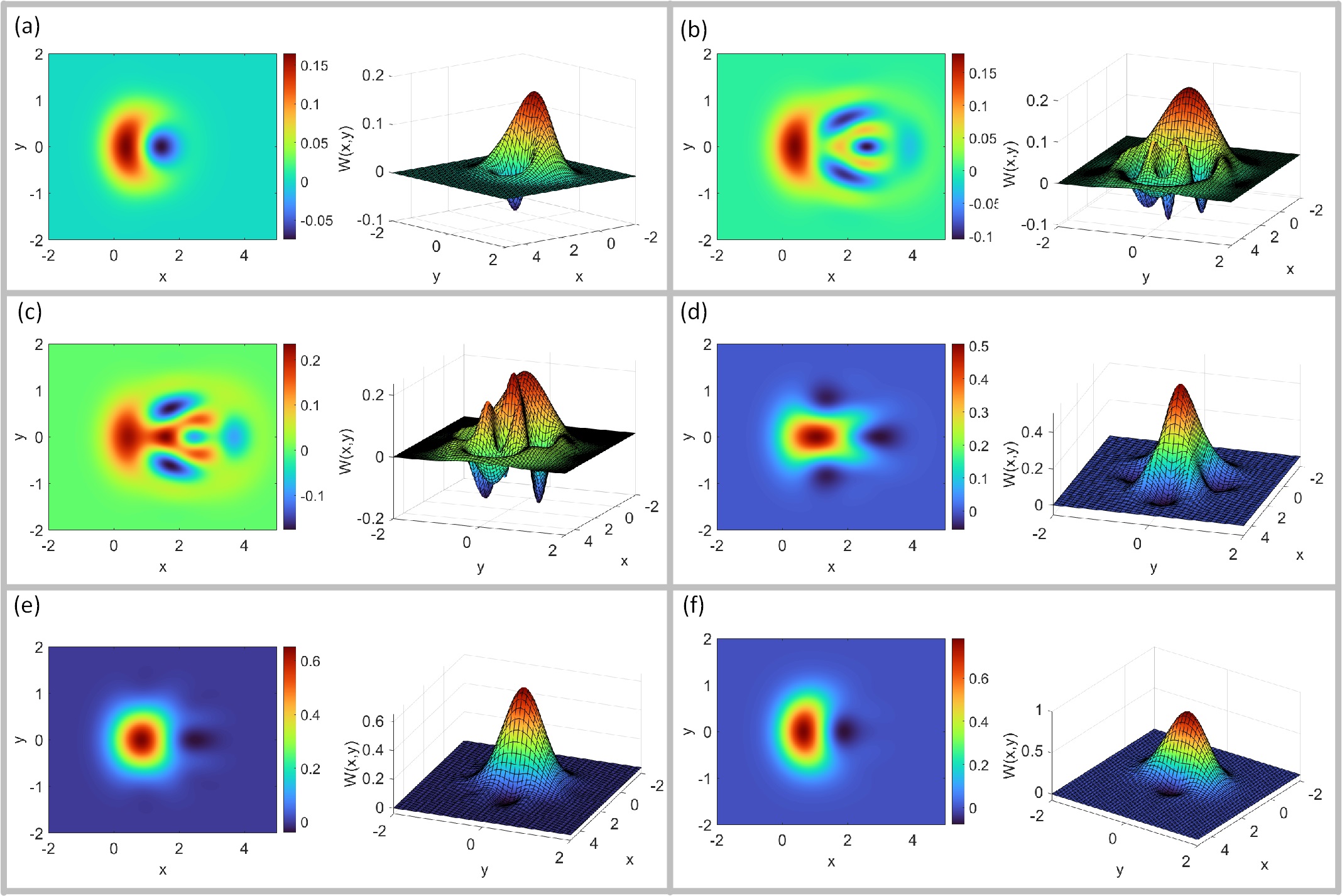}}
\caption{Wigner function for $\left| \alpha_{cat,s}  \right\rangle =\left( \left| {{\alpha }_{1}} \right\rangle -{{\zeta }_{1}}\left| {{\alpha }_{0}} \right\rangle  \right)+ \left( \left| {{\alpha }_{2}} \right\rangle -{{\zeta }_{2}}\left| {{\alpha }_{00}} \right\rangle  \right)$ with ${{\alpha }_{0}}=4,{{\alpha }_{00}}=2,\ \delta {{\alpha }_{00}}=-1.5$, and different values of $\delta {{\alpha }_{0}}=0$ (a), $\delta {{\alpha }_{0}}=-0.5$ (b), $\delta {{\alpha }_{0}}=-1$ (c), $\delta {{\alpha }_{0}}=-2$ (d), $\delta {{\alpha }_{0}}=-2.5$, $\delta {{\alpha }_{0}}=-3$}
\end{figure}

For $\delta {{\alpha }_{0}}=0$, the state $\left| {{\alpha }_{1}} \right\rangle -{{\zeta }_{1}}\left| {{\alpha }_{0}} \right\rangle =0$, and the Winger function is similar to a single cat state. Nevertheless, it varies significantly as ${{\delta\alpha }_{0}}$ increases, until about ${{\delta\alpha }_{0}}=3$, at which point the Winger function resembles a single cat state once more.   
\section{Conclusion}
In conclusion, in strong light-matter interaction through the HHG process, we may witness a considerable change in the quantum character of light by a small variation in parameter space. The Wigner function for superposition of two distinct optical Schrödinger cat states shows dramatic change around certain parameters. Our research helps develop quantum technologies like quantum sensing and opens the door to the engineering of new states of light. 
\section*{ACKNOWLEDGMENTS}
\nocite{*}
\section*{Appendix}
For the state $\left| {{\alpha }_{cat,s}} \right\rangle=\left| {{\alpha }_{1}} \right\rangle -{{\zeta }_{1}}\left| {{\alpha }_{0}} \right\rangle +\left| {{\alpha }_{2}} \right\rangle -{{\zeta }_{2}}\left| {{\alpha }_{00}} \right\rangle$
, the Wigner function can be written as
\begin{align}
  & W(\alpha )=\frac{1}{{{\pi }^{2}}}\frac{1}{N}\int{{{d}^{2}}\beta }\left( \left\langle  {{\alpha }_{1}} \right|-\zeta _{1}^{*}\left\langle  {{\alpha }_{0}} \right|+\left\langle  {{\alpha }_{2}} \right|-\zeta _{2}^{*}\left\langle  {{\alpha }_{00}} \right| \right) \nonumber\\ 
 & \quad \quad \quad \quad \hat{D}(\beta )\left( \left| {{\alpha }_{1}} \right\rangle -{{\zeta }_{1}}\left| {{\alpha }_{0}} \right\rangle +\left| {{\alpha }_{2}} \right\rangle -{{\zeta }_{2}}\left| {{\alpha }_{00}} \right\rangle  \right).{{e}^{-\beta {{\alpha }^{*}}+{{\beta }^{*}}\alpha }}  
\end{align}
, where $\hat{D}$ is displacement operator, $\alpha =x+iy,\quad \beta =u+iv$, and $N$ is normalization factor.
We have
\begin{align}
  & \left( \left\langle  {{\alpha }_{1}} \right|-\zeta _{1}^{*}\left\langle  {{\alpha }_{0}} \right|+\left\langle  {{\alpha }_{2}} \right|-\zeta _{2}^{*}\left\langle  {{\alpha }_{00}} \right| \right) 
 \hat{D}(\beta )\left( \left| {{\alpha }_{1}} \right\rangle -{{\zeta }_{1}}\left| {{\alpha }_{0}} \right\rangle +\left| {{\alpha }_{2}} \right\rangle -{{\zeta }_{2}}\left| {{\alpha }_{00}} \right\rangle  \right) \nonumber\\ 
 & =\left\langle  {{\alpha }_{1}} \right|\hat{D}(\beta )\left| {{\alpha }_{1}} \right\rangle -{{\zeta }_{1}}\left\langle  {{\alpha }_{1}} \right|\hat{D}(\beta )\left| {{\alpha }_{0}} \right\rangle 
  +\left\langle  {{\alpha }_{1}} \right|\hat{D}(\beta )\left| {{\alpha }_{2}} \right\rangle -{{\zeta }_{2}}\left\langle  {{\alpha }_{1}} \right|\hat{D}(\beta )\left| {{\alpha }_{00}} \right\rangle  \nonumber\\ 
 & -\zeta _{1}^{*}\left\langle  {{\alpha }_{0}} \right|\hat{D}(\beta )\left| {{\alpha }_{1}} \right\rangle +{{\left| {{\zeta }_{1}} \right|}^{2}}\left\langle  {{\alpha }_{0}} \right|\hat{D}(\beta )\left| {{\alpha }_{0}} \right\rangle  
  -\zeta _{1}^{*}\left\langle  {{\alpha }_{0}} \right|\hat{D}(\beta )\left| {{\alpha }_{2}} \right\rangle +\zeta _{1}^{*}{{\zeta }_{2}}\left\langle  {{\alpha }_{0}} \right|\hat{D}(\beta )\left| {{\alpha }_{00}} \right\rangle  \nonumber\\ 
 & +\left\langle  {{\alpha }_{2}} \right|\hat{D}(\beta )\left| {{\alpha }_{1}} \right\rangle -{{\zeta }_{1}}\left\langle  {{\alpha }_{2}} \right|\hat{D}(\beta )\left| {{\alpha }_{0}} \right\rangle   
  +\left\langle  {{\alpha }_{2}} \right|\hat{D}(\beta )\left| {{\alpha }_{2}} \right\rangle -{{\zeta }_{2}}\left\langle  {{\alpha }_{2}} \right|\hat{D}(\beta )\left| {{\alpha }_{00}} \right\rangle  \nonumber\\ 
 & -\zeta _{2}^{*}\left\langle  {{\alpha }_{00}} \right|\hat{D}(\beta )\left| {{\alpha }_{1}} \right\rangle +\zeta _{2}^{*}{{\zeta }_{1}}\left\langle  {{\alpha }_{00}} \right|\hat{D}(\beta )\left| {{\alpha }_{0}} \right\rangle 
  -\zeta _{2}^{*}\left\langle  {{\alpha }_{00}} \right|\hat{D}(\beta )\left| {{\alpha }_{2}} \right\rangle +{{\left| {{\zeta }_{2}} \right|}^{2}}\left\langle  {{\alpha }_{00}} \right|\hat{D}(\beta )\left| {{\alpha }_{00}} \right\rangle  
\end{align}
We may expand the terms in eq. (5) by using \cite{16,17}
\begin{equation}
\hat{D}(\alpha )\left| \beta  \right\rangle ={{e}^{(\alpha {{\beta }^{*}}-{{\alpha }^{*}}\beta )/2}}\left| \alpha +\beta  \right\rangle
\end{equation}
, and
\begin{equation}
\left\langle  \beta  | \alpha  \right\rangle ={{e}^{-\frac{1}{2}|\alpha {{|}^{2}}-\frac{1}{2}|\beta {{|}^{2}}+{{\beta }^{*}}\alpha }}
\end{equation}
Equation (4) can be used to evaluate the Wigner function by applying eqs. (5)–(7). In the integral calculation we may use the identity 
\begin{equation}
\int{{{d}^{2}}\beta {{e}^{-z|\beta {{|}^{2}}}}{{e}^{\beta x+{{\beta }^{*}}y}}=\frac{\pi }{z}{{e}^{xy/z}}}
\end{equation}
, which gives
\begin{align}
  & \pi^2 N.W(\alpha )={{e}^{2(-{{\alpha }_{1}}+\alpha )(\alpha _{1}^{*}-{{\alpha }^{*}})}}  
 -{{\zeta }_{1}}{{e}^{-\frac{1}{2}|{{\alpha }_{1}}{{|}^{2}}}}{{e}^{-\frac{1}{2}|{{\alpha }_{0}}{{|}^{2}}}}{{e}^{+\alpha _{1}^{*}{{\alpha }_{0}}}}{{e}^{2(-{{\alpha }_{0}}+\alpha )(\alpha _{1}^{*}-{{\alpha }^{*}})}} \nonumber\\ 
 & +{{e}^{-\frac{1}{2}|{{\alpha }_{1}}{{|}^{2}}}}{{e}^{-\frac{1}{2}|{{\alpha }_{2}}{{|}^{2}}}}{{e}^{+\alpha _{1}^{*}{{\alpha }_{2}}}}{{e}^{2(-{{\alpha }_{2}}+\alpha )(\alpha _{1}^{*}-{{\alpha }^{*}})}}
 -{{\zeta }_{2}}{{e}^{-\frac{1}{2}|{{\alpha }_{1}}{{|}^{2}}}}{{e}^{-\frac{1}{2}|{{\alpha }_{00}}{{|}^{2}}}}{{e}^{+\alpha _{1}^{*}{{\alpha }_{00}}}}{{e}^{2(-{{\alpha }_{00}}+\alpha )(\alpha _{1}^{*}-{{\alpha }^{*}})}} \nonumber\\ 
 & -\zeta _{1}^{*}{{e}^{-\frac{1}{2}|{{\alpha }_{1}}{{|}^{2}}}}{{e}^{-\frac{1}{2}|{{\alpha }_{0}}{{|}^{2}}}}{{e}^{+\alpha _{0}^{*}{{\alpha }_{1}}}}{{e}^{2(-{{\alpha }_{1}}+\alpha )(\alpha _{0}^{*}-{{\alpha }^{*}})}} 
 +{{\left| {{\zeta }_{1}} \right|}^{2}}{{e}^{2(-{{\alpha }_{0}}+\alpha )(\alpha _{0}^{*}-{{\alpha }^{*}})}} \nonumber\\ 
 & -\zeta _{1}^{*}{{e}^{-\frac{1}{2}|{{\alpha }_{2}}{{|}^{2}}}}{{e}^{-\frac{1}{2}|{{\alpha }_{0}}{{|}^{2}}}}{{e}^{+\alpha _{0}^{*}{{\alpha }_{2}}}}{{e}^{2(-{{\alpha }_{2}}+\alpha )(\alpha _{0}^{*}-{{\alpha }^{*}})}} 
 +\zeta _{1}^{*}{{\zeta }_{2}}{{e}^{-\frac{1}{2}|{{\alpha }_{0}}{{|}^{2}}}}{{e}^{-\frac{1}{2}|{{\alpha }_{00}}{{|}^{2}}}}{{e}^{+\alpha _{0}^{*}{{\alpha }_{00}}}}{{e}^{2(-{{\alpha }_{00}}+\alpha )(\alpha _{0}^{*}-{{\alpha }^{*}})}} \nonumber\\ 
 & +{{e}^{-\frac{1}{2}|{{\alpha }_{2}}{{|}^{2}}}}{{e}^{-\frac{1}{2}|{{\alpha }_{1}}{{|}^{2}}}}{{e}^{+\alpha _{2}^{*}{{\alpha }_{1}}}}{{e}^{2(-{{\alpha }_{1}}+\alpha )(\alpha _{2}^{*}-{{\alpha }^{*}})}}
 -{{\zeta }_{1}}{{e}^{-\frac{1}{2}|{{\alpha }_{2}}{{|}^{2}}}}{{e}^{-\frac{1}{2}|{{\alpha }_{0}}{{|}^{2}}}}{{e}^{+\alpha _{2}^{*}{{\alpha }_{0}}}}{{e}^{2(-{{\alpha }_{0}}+\alpha )(\alpha _{2}^{*}-{{\alpha }^{*}})}} \nonumber\\ 
 & +{{e}^{2(-{{\alpha }_{2}}+\alpha )(\alpha _{2}^{*}-{{\alpha }^{*}})}} 
 -{{\zeta }_{2}}{{e}^{-\frac{1}{2}|{{\alpha }_{2}}{{|}^{2}}}}{{e}^{-\frac{1}{2}|{{\alpha }_{00}}{{|}^{2}}}}{{e}^{+\alpha _{2}^{*}{{\alpha }_{00}}}}{{e}^{2(-{{\alpha }_{00}}+\alpha )(\alpha _{2}^{*}-{{\alpha }^{*}})}} \nonumber\\ 
 & -\zeta _{2}^{*}{{e}^{-\frac{1}{2}|{{\alpha }_{1}}{{|}^{2}}}}{{e}^{-\frac{1}{2}|{{\alpha }_{00}}{{|}^{2}}}}{{e}^{+\alpha _{00}^{*}{{\alpha }_{1}}}}{{e}^{2(-{{\alpha }_{1}}+\alpha )(\alpha _{00}^{*}-{{\alpha }^{*}})}}
 +\zeta _{2}^{*}{{\zeta }_{1}}{{e}^{-\frac{1}{2}|{{\alpha }_{0}}{{|}^{2}}}}{{e}^{-\frac{1}{2}|{{\alpha }_{00}}{{|}^{2}}}}{{e}^{+\alpha _{00}^{*}{{\alpha }_{0}}}}{{e}^{2(-{{\alpha }_{0}}+\alpha )(\alpha _{00}^{*}-{{\alpha }^{*}})}} \nonumber\\ 
 & -\zeta _{2}^{*}{{e}^{-\frac{1}{2}|{{\alpha }_{2}}{{|}^{2}}}}{{e}^{-\frac{1}{2}|{{\alpha }_{00}}{{|}^{2}}}}{{e}^{+\alpha _{00}^{*}{{\alpha }_{2}}}}{{e}^{2(-{{\alpha }_{2}}+\alpha )(\alpha _{00}^{*}-{{\alpha }^{*}})}}
 +{{\left| {{\zeta }_{2}} \right|}^{2}}{{e}^{2(-{{\alpha }_{00}}+\alpha )(\alpha _{00}^{*}-{{\alpha }^{*}})}} 
\end{align}
To calculate normalization factor, $N$, we consider
\begin{equation}
\left| {{\phi }_{cat,s}} \right\rangle =\frac{1}{\sqrt{N}}\left| {{\alpha }_{cat,s}} \right\rangle
\end{equation}
, where 
\begin{equation}
\left\langle {{\phi }_{cat,s}} \right.\left| {{\phi }_{cat,s}} \right\rangle =\frac{1}{N}\left\langle {{\alpha }_{cat,s}} \right.\left| {{\alpha }_{cat,s}} \right\rangle =1
\end{equation}
\bibliography{sample}

\end{document}